# Neurophysiology of gaze orientation: Core neuronal networks

Laurent Goffart[1,2], Julie Quinet[3] and Clara Bourrelly[4]

1 : Aix Marseille Université, CNRS, Institut de Neurosciences de la Timone, Marseille, France
2 : Aix Marseille Université, CNRS, Centre Gilles Gaston Granger, Aix-en-Provence, France
3 : Department of Optometry and Vision Science, The University of Alabama at Birmingham, Birmingham, USA
4 : University of Pittsburgh, Center for the Neural Basis of Cognition, Pittsburgh, USA

## Abstract

The appearance of an object triggers a shift of gaze toward its location. This orienting response consists of a rapid rotation of the eyes, the saccade, sometimes accompanied by a head rotation. In this chapter, instead of describing the path leading from the target-evoked retinal activity to the changes in muscle tension, we shall take the reverse path. Starting from the muscle contractions, we shall proceed upstream and describe the core neuronal networks in the brainstem and cerebellum that enable us to rapidly and accurately orient the foveae towards visual targets located at different eccentricities and depths.

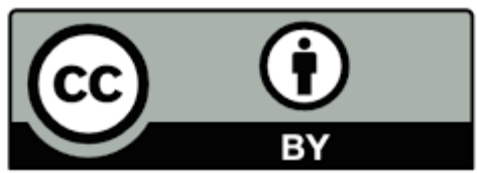

# Key points/objectives

Complementing several previous reviews, this chapter brings a synthesis of the knowledge that neurophysiologists and neuroanatomists gathered in the monkey during the last six decades, with the aim to explain the neuronal underpinnings of one of the most basic spatial abilities, i.e., directing gaze toward the location where something just happened.

# Introduction

Wrapped within the orbits by fibrous fascia and fat cushions, the eyeballs owe their orientation to the tensions that six extraocular muscles (EOM) exert upon them. Their steady tension acts as a kind of basal tone from which the eyes change their orientation once the balance of activity carried by their motor innervation also changes. Whenever modifications in the firing rate of motor neurons cause asymmetrical changes in the contraction of antagonistic extraocular muscles, the eyes start to move (Goffart 2009).

The figure 1A shows the approximate location of extra-ocular muscles. Four rectus muscles attach to the sclera anteriorly with respect to the eyeball equator whereas the oblique muscles insert posteriorly. The insertions of the lateral (LR) and medial (MR) rectus muscles are located on opposite sides of each eyeball, making their primary actions antagonistic with each other. Although the eyeballs do not rotate about rigid axes, by convention, their movements are described as rotations about three virtual axes: a vertical axis for horizontal (leftward and rightward) movements, a horizontal axis for vertical (upward and downward) movements and an anteroposterior axis for torsional movements (incyclotorsion and excyclotorsion). Kinematic studies in healthy subjects tested with the head upright measured the orientation of the eyeballs during fixation, saccades and pursuit eye movements and found that one single rotation axis suffices to change the orientation of each eyeball. When the eye is in primary position, the rotation axis lies within a plane called Listing's plane, which is perpendicular to the line of sight (Wong 2004). However, with training, subjects can make voluntary torsional eye movements (Balliet and Nakayama 1978). In most cases, this kind of movements is observed in case of lesions (Crawford and Vilis 1992; Helmchen et al., 1998) or when healthy subjects tilt their head (Tweed et al., 1995).

**< Figure 1 near here >**

With the head upright and the gaze directed straight ahead, the contraction of LR fibers and the relaxation of MR fibers causes the abduction of the eyeball while the contraction of MR fibers and the relaxation of LR fibers of the other eye causes its adduction. If the tensions exerted by the four other muscles do not change, gaze direction is expected to move only along the plane passing through the LR and MR muscles. For a long time, these muscles were considered to exert no secondary action, but the proposal of compartmentalization of extraocular muscles may lead to revise this view (Demer, 2015). Moreover, precise measurements made in human subjects revealed a transient torsion during horizontal saccades (Straumann et al., 1995), suggesting brief changes in the activation of the other extraocular muscles.

The neural control underlying the generation of vertical movements is more complicated. These movements involve not only the steady tension of LR and MR muscles, but also a complex cooperation between the four other muscles. The contraction of the superior rectus muscles (SR) supraducts (or elevates) the eyes whereas the contraction of the inferior rectus (IR) muscles infraducts (or depresses) them. These actions are stronger when the eye is in abduction. The contraction of oblique muscles also support vertical movements in addition to cause torsional movements. Their supraducting and infraducting actions are stronger when the eye is adducted. If we consider the movements made from the eyes centered in the orbit, then in addition to elevating the optic axis (cyan arrows in Fig. 1B), the contraction of the superior rectus (SR) muscle fibers causes adduction (black arrows) and incycloduction (or incyclotorsion; blue arrows) of the eyeball. Likewise, in addition to lowering the optic axis (yellow arrows in Fig. 1B), the contraction of the inferior rectus (IR) muscle fibers causes adduction and excycloduction (or excyclotorsion; green arrows). These secondary actions can be counteracted by the concomitant contraction of oblique muscles, for the contraction of inferior oblique (IO) muscle fibers causes elevation, abduction (red) and excycloduction (green), whereas the contraction of superior oblique (SO) muscle fibers causes depression, abduction and incycloduction (blue) (Demer 2019). Thus, while looking upward, the adduction and incycloduction caused by the contraction of the SR muscle fibers oppose to the abduction and excycloduction caused by the contraction of the IO muscles fibers. While looking downward, the adduction and excycloduction caused by the contraction of the IR muscle fibers oppose to the abduction and incycloduction caused by the contraction of the SO muscles fibers. In both cases, we shall see that the cancellation of adductive (or abductive) and cycloductive actions rests upon the muscle innervation by motor neurons located in the left and right parts of the midbrain. The bilateral innervation also accounts for the binocular aspect of eye movements.

# Binocular conjugate movements

Binocular conjugate movements are made when the head rotates while a visual target is being fixated (vestibulo-ocular reflex), when the whole visual scene is moving (optokinetic reflex) and whenever a target appears or is imagined in the peripheral visual field (saccades). In the following sections, we shall first examine the neurophysiology of saccade generation when the head does not move. Then, we shall study the case when a rotation of the head accompanies the saccade. During these combined movements of the eyes and head, two couplings can happen: either the eyes and the head turn in the same direction or they turn in opposite directions. Next, we shall see the neurophysiology of disconjugate movements of the eyes, which occur when gaze is shifted between targets located at different depths (vergence eye movements). Finally, we shall have an overview of the complex network that enables one to fixate and track a continuously moving target (smooth pursuit eye movements).

## Horizontal saccades

Figure 2 schematizes the current view of the premotor network generating horizontal saccades. Its neurons are located in the pontine and mesencephalic reticular formation. During leftward saccades (Fig. 2A), the fast-twitch muscle fibers of the right eye's MR and of the left eye's LR contract while those of antagonist muscles (right eye's LR and left eye's MR) relax. Motor neurons in the abducens nucleus (ABD, located in the pontine reticular formation) and in the oculomotor nucleus (OMN in the midbrain tegmentum) emit bursts of action potentials that cause the phasic contraction of fibers in the LR and MR muscles, respectively (Fuchs and Luschei 1970; Fuchs et al. 1988). The burst discharge of cells in the OMN is supported by spikes emitted by internuclear neurons (AIN) located in the contralateral abducens nucleus. The bursts of abducens motor (MN) and internuclear (AIN) neurons are driven by afferent input from excitatory burst neurons (EBN) located in the ipsilateral paramedian pontine reticular formation (PPRF, synapses a) (Keller, 1974; Luschei and Fuchs, 1972; Strassman et al., 1986a). The abducens neurons also receive excitatory input from internuclear burst-tonic neurons in the contralateral oculomotor nucleus (not shown). Discharging during saccades and increasing their firing rate during convergence, these cells are also involved in the motor drive for conjugate saccades. Indeed, in addition to cause exophoria, lidocaine injection in the medial rectus subdivision of the oculomotor nucleus causes hypometria and slowing of abducting saccades of the contralateral eye (Clendaniel and Mays, 1994).

Otherwise silent, the EBN fire bursts of action potentials whose interspike interval is smaller for larger saccades. For each saccade, the interspike interval is much more constant than expected if the instantaneous firing rate of EBN is not a replica of instantaneous eye velocity (Hu et al., 2007). Likewise, the suggestion that the firing rate of motoneurons would also "encode" the instantaneous eye velocity is questioned by a study in the cat (Davis-López de Carrizosa et al., 2011), which showed that a model factoring in muscle tension and its first derivative accounted for the firing rate of motor neurons better than a model factoring in eye position and its first and second derivatives (velocity and acceleration).

Regarding the relaxation of antagonist muscles, it results from a pause in the firing of motor neurons innervating them. This pause is itself caused by inhibitory input from burst neurons (IBN) located in the contralateral dorsal paragigantocellular reticular formation (dPGRF, synapses b). In addition to bursting during ipsiversive horizontal saccades (so called "on-direction" response), approximately half of these inhibitory burst neurons emit spikes also during contralateral horizontal saccades (off-direction response) (Kojima et al, 2008; Scudder et al., 1988; Strassman et al., 1986b). Thus, the agonist motor drive is a combination of excitatory and inhibitory presynaptic input from the ipsilateral EBN and contralateral IBN, respectively (van Gisbergen et al., 1981). The relative strength of these antagonist premotor inputs is under the descending influence of saccade-related burst neurons located in regions involved in other functions, such as the left and right caudal fastigial nuclei (located in the deep medial cerebellum) and the contralateral deep superior colliculus (in the midbrain tectum). In the monkey, the collicular input is disynaptic and the intermediate neurons have not been identified yet (Keller et al., 2000). The fastigial input, which is monosynaptic (Scudder et al., 2000) and bilateral (Noda et al., 1990), seems to exert an influence that is more modulatory than the collicular influence because, contrary to lesions in the superior colliculus (Goffart et al., 2012; Hikosaka and Wurtz, 1985; Sparks et al., 1990), fastigial lesions do not suppress or delay the generation of saccades (Bourrelly et al., 2018a). The contributions of the deep superior colliculi and caudal fastigial nuclei also differ insofar as horizontal saccades involve bursting activity of cells in only one superior colliculus, on the contralateral side

(Sparks et al., 1976), whereas saccade-related neurons burst in both (left and right) caudal fastigial nuclei (Fuchs et al., 1993; Ohtsuka and Noda, 1991).

< Figure 2 near here >

During rightward saccades (Fig. 2B), the active circuit is mirror reversed. The agonist muscles are the right eye's LR and the left eye's MR, whereas the antagonist muscles are the right eye's MR and the left eye's LR. Inhibitory burst neurons in the right dPGRF cause the relaxation of antagonist muscles. Under the excitatory influence of EBN in the right PPRF and the inhibitory influence of IBN in the left dPGRF, motor and internuclear neurons in the right abducens nucleus provide the motor commands to the ipsilateral LR and contralateral MR, respectively. During each saccade, the omnipause neurons (OPN) exhibit a pause in their otherwise sustained firing rate. The inhibition that they exert upon the EBN and IBN located in the left and right side of the PPRF resumes at saccade end.

Contrary to what the schematic figure 2 might suggest, the circuits producing rightward and leftward saccades are not completely symmetric insofar as we should not expect that equal number of active neurons, equal number of synapses and neuromuscular junctions. Likewise, the networks involved in the generation of downward and upward eye movements are not symmetrical either with respect to each other. In that case, we shall see that the connectivity involved in the generation of upward saccades is not the mirror image of the connectivity involved in the generation of downward saccades.

## Vertical saccades

The neuronal network underlying the generation of vertical saccades is located in the midbrain tegmentum (Büttner and Büttner-Ennever, 2006). For upward saccades (Fig. 3A), the agonist muscles are the SR and the IO and the antagonist muscles the IR and the SO. As reported above, in addition to elevating the eye, the contraction of SR muscle fibers incycloducts the eye, whereas the contraction of IO muscle fibers excycloducts it. Thus, the incycloduction caused by the motor commands to the SR combines with the excycloduction caused by the commands to the IO. The observation of negligible torsional component during upward saccades suggests that the two torsions cancel each other out. Such a balance of force likely involves bilateral activity in the midbrain because for each eye, the motor neurons innervating the SR muscle are located in the contralateral OMN whereas the motor neurons innervating the IO muscle are located in the ipsilateral OMN (Porter et al., 1983). During upward saccades, the two sets of commands must be adjusted in a timely manner for any asymmetry in the bilateral motor drive would engender a torsional component.

< Figure 3 near here >

The excitatory commands responsible for generating upward saccades are issued by burst neurons located in the left and right rostral interstitial nuclei of the medial longitudinal fasciculus (RIMLF) (Büttner et al., 1977). These commands are doubly bilateral (Fig. 3). Indeed, on the one hand, the bursts emitted by premotor neurons (uEBN) in the right RIMLF excite the motoneurons innervating the SR muscles of *both* eyes (purple synapses a) and the motor neurons innervating the IO muscles of *both* eyes as well (purple synapses b) (Moschovakis et al., 1991a). On the other hand, the burst emitted by uEBNs in the opposite (left) RIMLF excite the same motoneuronal groups (blue synapses a and b). Moreover, in addition to recruiting the motor neurons, the uEBN also excite neurons that inhibit the motor neurons innervating the antagonist muscles (IR and SO, synapses c). Located in the interstitial nucleus of Cajal, these cells (uIBN) inhibit the motor neurons in the contralateral oculomotor nucleus, i.e., cells responsible for the contraction of the IR of the contralateral eye (synapses d) and cells in the trochlear nucleus responsible for the contraction of the SO of the ipsilateral eye (synapses e). However, contrary to uEBN, uIBN do not send bilateral projections to the motor neurons innervating the SO muscle fibers. Their projections make synaptic contacts with motor neurons on the contralateral side only (Horn et al., 2003).

As for the upward saccades, the discharge of neurons located in each side of the brain generates a downward movement of *both* eyes (Fig. 3B). The premotor excitatory burst neurons responsible for downward saccades (dEBN) are located in the RIMLF, like the uEBN (Moschovakis et al., 1991b). However, unlike them, the dEBN do not send bilateral projections to the motoneurons; their projection is ipsilateral only. During downward saccades, the SO and IR muscle fibers of *both* eyes contract because the dEBN excite 1) the motor neurons in the ipsilateral OMN, which support the contraction of IR muscle fibers of the *ipsilateral* eye (synapses a), and 2) the motor neurons which, in the ipsilateral trochlear nucleus, support the contraction of SO muscle fibers of the *contralateral* eye (synapses b). In addition to lowering the eyes, the contraction of SO muscle fibers causes an incyclotorsion of the eye while the contraction of IR muscle fibers causes its excyclotorsion. Thus, the excyclotorsion caused by spikes from the ipsilateral OMN combine with the incyclotorsion caused by spikes from the contralateral trochlear nucleus. Precisely how these opposite torsional components cancel each other out during downward saccades remains unexplained. A bilateral adjustment of premotor commands seems inevitable. Concerning the neural processes in charge of the relaxation of antagonist muscles during downward saccades, they remain to be characterized.

Even though the number of cells and the number of synaptic boutons likely differ between the left and right sides of the reticular formation, the connectivity patterns involved in generating leftward and rightward eye movements are symmetrical with respect to the brainstem midsagittal plane (see Fig. 2 and 6). By contrast, the networks involved in the generation of downward and upward eye movements are not symmetrical (Fig 3).

During vertical saccades, the EBN involved in the generation of horizontal saccades do not fire. By contrast, some IBN in the dPGRF emit action potentials whose number increases with the saccade size (Scudder et al., 1988). For vertical saccades to remain straight while their size increases, one possibility could have been that the motor and internuclear cells in the abducens nucleus emit additional spikes, thus reducing the contrapulsive effect of IBN spikes, but this does not seem to be the case: MN and AIN in the abducens nucleus do not emit action potentials during vertical saccades. One possibility is that the contrapulsion exerted by IBN in the left and right dPGRF cancel each other out. If no perfect symmetry exists between the set of active neural elements (number of neurons and number of synaptic boutons) located on each part of the reticular formation, then the observation of strictly vertical saccades leads us to infer that a process located upstream of the IBN adjusts their bilateral activity during vertical saccades. We shall see that the saccade-related burst neurons in the caudal fastigial nucleus may account for this adjustment.

The bilateral control is not restricted to the group of IBNs involved in the generation of horizontal saccades. In addition to elevating both eyes, the discharge of motor neurons in the right OMN causes, as a secondary action and via the contraction of the right eye's IO and the left eye's SR, a rightward deflection of both eyes during vertical saccades while the discharge of motor neurons in the opposite OMN causes a leftward deflection (Fig. 3A). Likewise, in addition to lowering both eyes, the dEBN in the right RIMLF lead to the contraction of the left eye's SO and the right eye's IR, causing a leftward deflection of their trajectory while dEBN in the left RIMLF cause a rightward deflection during vertical saccades (Fig. 3B, top row). Thus, the production of vertical saccades that are straight (with no horizontal deviation) implies that the horizontal deflections caused by the secondary action of the muscles elevating or depressing the eyes are also canceled in the motor commands.

Inn summary, the generation of straight vertical saccades rests upon a balance of bilateral motor activation, which itself depends upon an adjustment of premotor commands that takes into account the structural (neurons and their connectivity) and functional (firing properties, secondary and tertiary consequence of muscle contractions) asymmetries between the oculomotor territories that are distributed on either side of the brain stem. It is possible that the saccade-related neurons in the left and right fastigial nuclei and in the oculomotor vermis of the cerebellum regulate this bilateral balance. Indeed, most of them burst during vertical saccades (Fuchs et al., 1993; Ohtsuka and Noda, 1991, 1995) whereas their asymmetrical functional perturbation horizontally deflects the trajectory of vertical saccades (Takagi et al. 1998; Goffart et al., 2003; Goffart et al., 2004; Nitta et al. 2008; Bourrelly et al., 2021). Bilateral activity in the two deep superior colliculi is also involved with their medial halves for driving upward saccades and their lateral halves driving downward saccades.

## Combined horizontal and vertical saccades

When a target appears off the cardinal visual meridians (horizontal meridian passing through the LR and MR muscles and midsagittal vertical meridian), the orienting response does not consist of horizontal and vertical saccades made in sequential order. Both horizontal and vertical components are generated in such a way that their onset and offset are relatively simultaneous. These combined horizontal-vertical saccades are conventionally called "oblique" saccades. Considering the physical diversity of muscles (lengths, contractile properties and orbital attachments) and the fact that their innervations are not located in nearby regions of the brainstem, the simultaneous onsets of the horizontal and vertical components and the limited curvature of the saccade when gaze is shifted between two targets located at the same distance are empirical observations suggesting the existence ofmechanisms coupling their time course.

In the medial pontine reticular formation, the nucleus raphe interpositus (RIP) hosts a population of cells that exhibit a sustained firing rate during intersaccadic intervals but pauses during all saccades, regardless of their direction and amplitude (Keller 1974; Phillips et al., 1999; Strassman et al., 1987). Called omnidirectional pause (or omnipause) neurons (OPN), these cells are thought to prevent the eyes from moving during visual fixation through their inhibitory post-synaptic influence upon the premotor burst neurons (Fig. 2). However, the pause of their firing rate could also be responsible for synchronizing the onsets of horizontal and vertical saccades, as they project to burst neurons in the PPRF and RIMLF (Ohgaki et al., 1989). Up to now, experimental studies did not document onset asynchronies during the inactivation or the lesion of RIP. No fixation instability was reported either (Kaneko, 1996; Soetedjo et al., 2000). The absence of deficit in visual fixation may have resulted from limited testing conditions insofar as the consequence of their dysfunction was not tested during delayed or memory-guided saccade tasks. The observation of irrepressible saccades toward the peripheral target during the delay interval would unambiguously confirm the role of OPN in maintaining steady the gaze direction during visual fixation, as electrical microstimulation studies suggested (Gandhi and Sparks 2007; Keller 1974). However, these studies are not sufficiently conclusive because the interruption and the delaying of saccades caused by microstimulation might result from the (undesired) retrograde recruitment of inhibitory input to the pause neurons. A projection from inhibitory burst neurons to the pause neurons has indeed been reported in the cat (Takahashi et al. 2022). This projection remains to be confirmed in the monkey because the available data do not support it (Strassman et al. 1986b).

Turning back to the generation of combined horizontal and vertical saccades, one study reported crucial observations signaling a process that couples their time course during oblique saccades. When the horizontal component is experimentally slowed by the local injection of a pharmacological agent in the pontine reticular formation, the vertical component is also slowed (Barton et al., 2003; Sparks et al., 2002). In some of these experiments, the duration of both components was prolonged while the accuracy of saccades was preserved, indicating the persistence and preservation of the target-related commands. The neural connectivity and processes underlying this "component stretching mechanism" in primates have not yet been elucidated.

# Combined eye and head movements

< Figure 4 near here >

Saccades are frequently associated with a head movement when the target eccentricity approaches and exceeds the limits of the oculomotor range. Moving the head enables to look at targets located more eccentrically in the visual field, and to re-center the eyes in their orbit. Added together, the deviation of the eyes in the orbit and the deviation of the head relative to the trunk determine the direction of gaze. Figure 4 shows the recording of a combined horizontal movement of gaze and head while a monkey orients toward the location of a static visual target. The orientation of the eye in the orbit (blue trace) is obtained by subtracting the orientation of the head in space (red trace) from the orientation of gaze (green trace). Two functional modes characterize the time course of the eye and head movements. A mode during which both eyes and the head rotate in the same direction and a mode during which they move in opposite directions. The second mode happens later because the eyes move much faster than the head and because the restoration of the equilibrium of bilateral activity that determines gaze direction has been completed before the equilibrium that determines head direction. This mode also happens when the head movement precedes the eye saccade, mostly during cases in which the target location is known in advance (Bizzi et al. 1972). Before describing how gaze is being stabilized and ends its orienting movement, we shall examine the first mode, i.e., how despite the vestibulo-ocular reflex, the eyes and the head can move in the same direction as during horizontal gaze shifts.

## Combining a head movement with a saccadic eye movement

In this section, we limit our description to the horizontal eye movements that are made while the head also moves horizontally. Whether the head is free to move or not, small horizontal gaze movements are comparably accurate (Freedman and Sparks, 1997; Quinet and Goffart, 2003; Tomlinson and Bahra, 1986). However, large gaze shifts are more accurate when the head is free to move than when it is immobilized (Phillips et al., 1995). In the monkey, about twenty muscles are involved to change the orientation of the head (Richmond et al., 2001). Among them are the muscles whose contraction changes the orientation of the head relative to the trunk, the muscles that stabilize the head during its rotation and the muscles that do not exhibit any change (Lestienne et al., 1995). The co-contraction of the left and right *rectus capitis posterior minor* (RCPm) stabilize the atlas while the contraction of the *rectus capitis posterior major* (RCPM) and *obliquus capitis inferior* (OCI) produce an ipsilateral horizontal rotation of the head. The observation that the contralateral muscles exhibit no electromyographic change indicates that stopping the head rotation is not caused by muscle co-contraction (Lestienne et al., 1995). For movement amplitudes greater than 20 degrees, the contraction of the *splenius capitis* muscle adds to that of the RCPM and OCI muscles. At the end of the orienting movement, the co-contraction of the RCPm, RCPM, OCI and *obliquus capitis superior* (OCS) muscles altogether stabilizes the head (Corneil et al., 2001; Lestienne et al., 1995).

Thus, orienting gaze shifts mobilize several muscles when a movement of the head accompanies the saccade. The synergy results from activities distributed within a complex set of motor and premotor neurons located in the reticular and vestibular nuclei (Cullen 2009; Sparks et al., 2001). Contrary those involved in the cat (Takahashi and Shinoda, 2018), the premotor processes remain incompletely characterized in the monkey. Contemporary knowledge is not sufficient to reach an understanding of the neural control of combined eye-head gaze shifts comparable to the neurophysiology of saccades.

The recruitment of EBN and reticulo-spinal neurons (RSN) by descending commands from bursting cells in the deep superior colliculus (grey-colored synapses a in Fig. 5) leads to the excitation of the pools of extraocular motoneurons (ABD) in the abducens nucleus and neck motoneurons (NMN) in the cervical spinal cord (green and blue synapses b), and thus, to the contraction of the agonist extraocular and neck muscle fibers, respectively. Contrary to the cat, collicular connections with EBN do not seem to be monosynaptic in the monkey (Keller et al., 2000). The EBN and RSN also excite the IBN (synapse a'), thereby suppressing the activity of cells that otherwise act to facilitate the contraction of antagonist extraocular (synapses c) and neck muscles (synapse g), respectively. Finally, they also excite secondary vestibular neurons which are inhibitory (IVN-2; blue and green colored synapses d) and located in the ipsilateral medial vestibular nucleus. Within the vestibular nuclei, the discharge of IVN-2 inhibits the secondary vestibular neurons, IVN-1 and EVN-1 (purple-colored synapse e). By inhibiting IVN-1, the motor neurons innervating the agonist extraocular and neck muscles fibers (red synapses f and f') are disinhibited, and the contraction of these muscles is facilitated. By inhibiting the EVN-1, the IVN-2 prevent the primary vestibular neurons (PVN), whose firing is enhanced by the head movement, from causing a counter-rotation of the eyes in the orbits (synapse h). The contraction of antagonist neck muscles through vestibulospinal pathways is also prevented (synapse h').

< Figure 5 near here >

## Stopping gaze while the head terminates its motion

The second mode of eye and head movements is similar to the vestibulo-ocular reflex recorded when the head is passively rotated while subjects maintain their gaze directed toward a visual target. During active eye and head movements, the premotor input to motor neurons makes a transition from commands fed by target-related (visual or auditory) signals to commands fed by head movement-related signals (Cullen, 2009; Cullen and Roy, 2004). The latter consist of trains of action potentials originating in the three semicircular canals buried within the temporal bone. Each semicircular canal consists of a ring within which a fluid (endolymph) flows and mechanically deflects the stereocilia of multiple hair cells whenever the head rotation accelerates or decelerates. This deflection either increases or decreases the discharge of primary vestibular neurons (PVN) with a rate proportional to head velocity.

The two lateral canals are located in the left and right inner ears and lie roughly in a "plane" which is tilted not only relative to the Reid plane, but also to the "plane" passing through the medial and lateral rectus muscles (Cox and Jeffery, 2008; Ostriker et al., 1985). These canals are often called "horizontal" because they are primarily activated during head rotations about the Earth gravity axis. During such rotations, PVN on the side toward which the nose is rotating (ipsilateral side) increase their firing rate while those on the contralateral side decrease their firing rate. These bilateral changes modify the firing rate of post-synaptic neurons (secondary vestibular neurons) located in the medial and superior vestibular nuclei (McCrea et al., 1987b). On either side of the brain's midsagittal plane, these nuclei are bilaterally organized and house two categories of neurons: type 1 and type 2 neurons. Type 1 neurons increase their firing rate during ipsiversive head rotations, i.e., toward the side (left or right) containing the nucleus in which their discharge is recorded. During contraversive head rotations, they decrease their discharge. In addition, they pause during saccades (McCrea et al., 1987). In the monkey, these neurons have been called Position Vestibular Pause (PVP) neurons. Type 2 neurons exhibit the opposite pattern, increasing their firing rate during contraversive head rotations and decreasing it during ipsiversive head rotations (Fuchs et al., 2005; Roy and Cullen, 2002; 2003).

< Figure 6 near here >

Figure 6 schematizes the mechanism by which vestibular signals contribute to ending a leftward movement of gaze toward a visual target. The end of the orienting movement is characterized by a time interval during which the gaze direction remains stationary relative to the trunk while the head slowly terminates its motion. The saccades by both eyes give way to their rotation in the opposite direction. As the head rotates, the primary vestibular neurons (PVN) excite the type I vestibular neurons (EVN-1) in the ipsilateral medial and ventrolateral vestibular nuclei (Broussard et al., 1995). In turn, the EVN-1 initiate and drive the counter-rotation of both eyes in the orbit by exciting the motor (LRMN) and internuclear (AIN) neurons in the contralateral abducens nucleus (red-colored synapses a). Thus, the abducens neurons can resume their activity from the pause imposed by the inhibitory input from IBN in the left dPGRF (see blue-colored synapses b in Fig. 2). The LR muscle fibers of the right eye can contract, as well as the MR muscle fibers of the left eye, which are activated via motoneurons located in the OMN. The latter cells (MRMN) are also excited by neurons in the ipsilateral nucleus of the ascending tract of Deiters (ATDN; green-colored synapse b). In parallel, the spikes emitted by other type 1 vestibular neurons (IVN-1) choke the activity of the LRMN and AIN in the ipsilateral abducens nucleus (red-colored synapses c), preventing them from responding to any residual descending (target-related) commands. The recruitment of type 2 neurons (IVN-2) in the contralateral vestibular nuclei (red-colored synapse d) complements the cessation of agonist commands by inhibiting the activity of EVN-1, thereby reducing their excitatory influence upon the motor and internuclear neurons (synapses e). IVN-2 also inhibit the activity of IVN-1, which in turn favors the disinhibition of cells (synapses f) in charge of the counter-rotation of the eyes under the influence of EVN-1 (synapses a) (Broussard et al., 1995; Scudder and Fuchs, 1992).

The functional anatomy schematized in Fig. 6 enables us to see how the disinhibition of type 1 vestibular neurons can quickly stop the eye saccade during ipsilateral orienting movements. The spikes of IVN-1 (inhibitory synapses c) choke the burst commands while the spikes of EVN-1 order the eyes to counter-rotate as the head terminates its motion (excitatory synapses a). This mechanism also helps us to understand the hypometria of contralateral eye and head movements after unilateral lesion of the caudal fastigial nucleus (for instance the right nucleus) (Quinet and Goffart, 2007). Contralateral (leftward) saccades are hypometric because the agonist drive from EBN lacks of the excitatory input from inactivated (right) fastigial cells, and because the saccade-related burst of neurons in the opposite (unaffected) fastigial nucleus excite IBN in the contralateral (right) reticular formation. These IBN inhibit IVN-2 in the opposite (left) vestibular nuclei, disinhibiting the excitatory influence of EVN-1 (initiating the counter-rotation of the eyes) and the inhibitory influence of IVN-1 upon ipsilateral motor neurons (inhibiting the saccade-related drive). The latter mechanism also explains the hypometria of head movements. Their amplitude is truncated because the IVN-1 inhibits the ipsilateral neck motor neurons (NMN; synapse e) while the EVN-1 facilitates the contraction of contralateral neck muscles (synapse g).

< Figure 7 near here >

As for the generation of upward and downward saccades (Fig. 3), the neural connectivity involved in the generation of downward and upward slow eye movements are not symmetrical (McCrea et al., 1987a). Figure 7A schematizes the network that stabilizes the visual field at the end of a downward gaze movement. While the head pitches down and slowly terminates its motion, both eyes rotate upward in the orbits. The upward eye movement is the outcome of contracting SR and IO muscles fibers. Driven by the increased firing of PVN connected with the left anterior canal, excitatory vestibular neurons (u-EVN) in the left medial vestibular nucleus activate the motor neurons innervating the left eye's SR and the right eye's IO muscle fibers (synapses a).

In addition to elevating both eyes, the contraction of these fibers causes an incyclotorsion of the left eye and an excyclotorsion of the right eye, as if both eyes were counteracting a tilt of the head toward the left shoulder. This clockwise torsion of both eyes can be attenuated by the torsion caused by the excitatory input from u-EVN in the right MVN. Because of their excitation by the right anterior canal, PVN drive the firing rate of u-EVN, which in turn excite the motor neurons that innervate the right eye's SR and the left eye's IO muscle fibers (synapses b), causing the right eye's incyclotorsion and left eye's excyclotorsion. The absence of torsion during strictly upward slow eye movements suggests that the output of the left and right vestibulo-oculomotor channels have been adjusted so that the torsional signals balance out.

Regarding the antagonist muscles, their inhibition is promoted by u-IVN located in the superior vestibular nuclei. These cells inhibit the motor neurons innervating the IR of the ipsilateral eye (synapses c) and those innervating the SO muscle of the contralateral eye (synapse d). Contrary to the network involved in horizontal eye and head movements (Fig. 6), the neural mechanisms triggering the counter-rotation of the eyes when the head pitches down are not yet elucidated. However, the inputs are necessarily bilateral and balanced since the eyes move by the same amount and their torsion canceled.

Finally, figure 7B schematizes the neuronal network involved in stopping an upward gaze shift while the head slowly terminates its upward motion. Both eyes rotate downward in the orbits because of the contraction of IR and SO muscle fibers. This is driven by the increased firing of primary vestibular neurons connected with the left posterior canal. Excitatory vestibular neurons (d-EVN) in the left medial vestibular and ventral lateral vestibular nuclei activate the motor neurons innervating the fibers of the right eye's IR and the left eye's SO muscles (synapses a).

As above, in addition to depressing both eyes, the contraction of these fibers cause an excyclotorsion of the right eye and an incyclotorsion of the left eye, as if the eyes are counteracting a tilt of the head toward the right shoulder. Once again, the binocular torsion is attenuated by a torsion made in the opposite direction and caused by excitatory input from the right posterior canal to the motor neurons innervating the left eye's IR and the right eye's SO muscle fibers (synapses b). Inhibitory vestibular neurons (d-IVN) located in the superior vestibular nuclei prevent the antagonist muscles from contracting by inhibiting the motor neurons innervating the SR of the contralateral eye (synapses c) and possibly also the motor neurons innervating the IO muscle of the ipsilateral eye (synapse d). As for the neural mechanisms ending downward gaze shifts, the neural mechanisms triggering the counter-rotation of the eyes when the head pitches up are not completely identified. Nevertheless, if the eyes do not exhibit any torsional component, the motor commands are necessarily bilateral and balanced.

In summary, by studying the networks underlying the generation of slow eye movements, we discover a complexity that radically differs from the straightforward solutions that human engineers would have designed. One puzzling question concerns the constraints that led the torsional components caused by the contraction of elevator or depressor muscles to cancel each other out. Moreover, the connectivity that we have described is not complete, because new eye movements can be produced in response to "unusual" combinations of head rotations and visual stimulation. Experiments in the cat have shown that the visual stimulation exerts a significant influence on the functional connectivity yielding the vestibulo-oculomotor responses. Horizontal compensatory eye movements can be generated during vertical head rotations in the dark after training sessions during which the visual field moved horizontally whenever the head rotated vertically. The horizontal nystagmus persisted in the absence of any new visuo-vestibular experience (Schultheis and Robinson, 1981).

## Fixating a static visual target

Considering the line of sight as the direction of gaze measured when a subject fixates a visual target, the binocular fixation point designates the intersection between the two lines of sight (Collewijn et al., 1997). Of course, the quantification of movements must be distinguished from their neurophysiological correspondence. Looking at a visual target obviously does not reduce to the excitation of one single photoreceptor on the fovea of each eye. We shall therefore speak of a binocular fixation zone. On the motor side, the orientation of the eyes is maintained steady by the sustained discharge of a tonic neurons distributed in the nucleus prepositus hypoglossi (NPH), the vestibular nuclei (in the medullary reticular formation) and in the interstitial nucleus of Cajal (INC, Fukushima et al., 1992; McFarland and Fuchs, 1992; Moschovakis, 1997). The more deviated the eyes in the orbit, the higher their firing rate. In addition to connective tissues that wrap each eyeball within the orbit, two neuronal networks are involved in maintaining the rotation axes of the eyes within Listing's plane and deviating them horizontally and vertically. These networks are relatively independent from each other insofar as a lesion in the NPH primarily affects the ability to maintain gaze direction within the horizontal plane, with marginal effects on vertical gaze direction (Kaneko, 1997; 1999). After the saccade, the amplitude of which is unaffected, the eyes exhibit an exponential drift toward their primary orientation (roughly straight ahead). By contrast, lesion in the INC causes a failure to hold vertical deviations of the eyes (i.e. both upward and downward) but not horizontal deviations. A torsional drift of the eyes is also observed when the lesion is unilateral (Helmchen et al., 1998).

Considering now the case during which the foveation changes between targets or locations situated at different depths in the three-dimensional physical space, three visuomotor processes occur: 1) vergence, 2) accommodation and 3) variation of pupil diameter. Convergence designates the symmetrical eye movements made when the MR muscle fibers of both eyes contract, diminishing the distance of the binocular fixation zone relative to the head. Divergence designates those movements made when the distance increases because of a binocular relaxation of MR muscle fibers. The contraction level of intraocular muscles also changes (McDougal and Gamlin, 2015). During convergence, the contraction of ciliary muscles (CM) reduces the tension in zonular fibers, changing the refractive power of the lens as its anterior surface bulges forward (increase in convexity). Thus, the target images are brought into focus on the foveae. This accommodation is associated with a myosis, i.e., the constriction of the iris and the reduction of the pupil diameter, thereby increasing the depth of field. The myosis itself results from the contraction of the sphincter pupillae (SP) combined with the relaxation of the dilator pupillae (DP). Altogether, these three visuomotor processes constitute the so-called near response (or near triad).

Although these processes are generated jointly, clinical studies report pathological cases in which they are dissociated and the neuroanatomy helps us to understand why. Figure 8 shows that the contraction of MR muscle fibers is controlled by the firing rate of motor neurons (MRMN, in red) located in the oculomotor nucleus whereas the intrinsic musculature of each eye is controlled by motor neurons located in the ciliary ganglions (CG) and premotor neurons in the Edinger-Westphal preganglionic nucleus (EWpg) (Fig. 8, in black). In both cases, the innervation is ipsilateral (Gamlin, 1999). The binocular control and the near triad starts with their afferent input from two major regions of the midbrain tegmentum: the supraoculomotor area (SOA in magenta) and the central mesencephalic reticular formation (cMRF, in purple) (May et al., 2018; 2019).

**< Figure 8 near here >**

The SOA is primarily involved in the generation of the slow near response and its sustained maintenance. Therein, we find burst neurons and tonic neurons, with separate subgroups characterized by different discharge patterns during the generation of diverging and converging eye movements. More common just lateral to the MLF, four categories of neurons are found: convergence burst neurons, divergence burst neurons, convergence tonic neurons and divergence tonic neurons (Mays, 1984; Mays et al., 1986). Whether divergence neurons exert an inhibitory influence upon the firing rate of convergence neurons is not known yet. Neurons in SOA send projections to the medial rectus MN that are ipsilateral, whereas the projections to the EWpg are bilateral (May et al., 2018) (not shown).

In the cMRF, we find neurons that emit a burst discharge whenever the near response accompanies a saccadic eye movement: saccade convergence burst neurons and saccade divergence burst neurons. These saccade-vergence burst neurons exhibit a burst of spikes related to vergence velocity during either convergence saccades or divergence saccades. The former fire during asymmetrical saccades associated with an increase of the vergence angle and the latter fire during asymmetrical saccades associated with a divergence (Quinet et al., 2020). Each cMRF innervates the SOA and the EWpg bilaterally. It also sends ipsilateral projections to the sites in the OMN in which the motor neurons that innervate the singly-innervated muscle fibers are located (May et al., 2018) (not shown).

The bilateral projections from SOA to the EWpg, and from the cMRF to both EWpg and SOA are completed by bilateral afferents from caudal fastigial nuclei (cFN, in blue) (Bohlen et al. 2021; May et al. 1992) whose contribution in the near-triad has been reported (Bohlen et al., 2021; Gamlin and Zhang, 1996). In addition to send bilateral projections to the midbrain tegmentum and contralateral projections to the pontomedullary reticular formation (where EBN and IBN are located, Fig. 2), the cFN also project bilaterally to the rostral part of the deep superior colliculi (in green), with a contralateral predominance (May et al., 1990). Therein, some cells either increase (convergence neurons) or decrease (divergence neurons) their firing rate during vergence eye movements when recording in the SOA or deep cerebellar nuclei (cFN or Posterior Interposed nucleus) (Bohlen et al., 2021; Gamlin, 2002; Gamlin et al., 1996; Zhang and Gamlin, 1996; 1998).

In addition to modulations during vergence (van Horn et al., 2013), recording studies in the rostral superior colliculi also revealed the presence of cells that burst during miniature saccades (also called microsaccades), like more caudal cells discharge during larger saccades (Hafed and Krauzlis, 2012). Local pharmacological inactivation confirmed their causal role in the generation of microsaccades (Hafed et al., 2009). However, it also unraveled an implication of rostral SC in specifying the direction of gaze when a static or moving target is presented (or even imagined) in the central visual field (Hafed and Krauzlis, 2008; Hafed et al., 2008). Indeed, unbalanced activity between the two rostral SC causes a fixation offset: gaze direction is not directed anymore to the target center but toward an offset location. The same deficit happens after unilateral pharmacological perturbation of cFN, suggesting that the sustained firing rate of its neurons adjusts the balance of activity between the two rostral SC during visual fixation (Bourrelly et al., 2018b; Goffart et al., 2012; Guerrasio et al., 2010). Thus, the direction of gaze would correspond to an equilibrium of commands that counterbalance with each other, involving fastigial projections to the left and right rostral SC but also to the left and right nuclei *reticularis tegmenti pontis* (NRTP). Through their projections to NRTP, the two rostral SC and cFN have access to neurons which, in the flocculus and paraflocculus, influence the firing rate of tonic neurons that innervate the motor and internuclear neurons in the abducens nuclei (Belton and McCrea, 2000; Lisberger et al., 1994; Noda and Mikami, 1986; Noda and Suzuki, 1979; Noda et al., 1990).

In summary, the knowledge of the detailed neuronal circuitry underlying the generation of disjunctive eye movements is not advanced as much as the circuitry for conjugate saccades (Mays and Gamlin, 1995a, b) but a lot of progress has been made during the last decade (e.g., May et al., 2018, 2019, 2022; Quinet et al., 2020). Soon it will be possible to understand the complex neurophysiology that underlies the generation of vergence eye movements and their dysfunction (Das, 2016; Lennerstrand, 2007).

# Tracking a moving visual target

A target drifting in the visual field often elicits a primary saccade (called interceptive saccade) at the end of which gaze does *not* land ahead of the moving target. Interceptive saccades either accurately capture (foveate) or fall behind the target. Then, a slow eye movement follows with a velocity that rarely matches the target's. A textured background

(yielding visual motion in the direction opposite to the eye movement) does not seem to be responsible for the pursuit slower than target velocity because the consequences of its suppression are very weak (Keller and Khan, 1986). A subsequent correction saccade (called catch-up saccade) restores the foveation. Thus, unless the target trajectory is already known and foreseeable, gaze lags behind the target most of the time (Goffart et al., 2017a), especially when the target is small: the visual tracking is mostly composed of catch-up saccades interspersed by intervals during which the eyes drift roughly in the same direction as the target and with a lesser speed. In well-trained subjects, the catch-up saccades become less frequent (Botschko et al., 2018; Bourrelly et al., 2016). Moreover, increasing the target size enhances the velocity of slow eye movements and reduces the size and number of catch-up saccades (Pola and Wyatt, 1985). This finding is important because most experimental studies use a small spot of light as a target.

Contrary to a static target which yields a bounded activity, a moving spot evokes on the retina a streak of activity the tail of which corresponds to past locations of the target and the leading edge the most recent location. In the deep superior colliculus, as for saccades to a static target, the population of saccade-related burst neurons is expanded (Anderson et al., 1998; Goossens and van Opstal, 2006; Sparks et al., 1976). However, it does not recruit cells that fire during saccades to future target locations. The population consists of a continuum of cells ranging from neurons issuing commands related to past locations of the target to neurons issuing commands related to its current location (Goffart et al., 2017). At the level of motor neurons, other input must therefore complement the collicular signals. This input stems, at least in part, from motion signals processed in area MT because its lesion leads to saccades landing at locations that the target crossed earlier. Thus, the ability to foveate the visual target may consist of adding a "predictive" component based on the integration of putative target-related "velocity" signals to collicular signals (Keller et al., 1996; Optican, 2009). However, an alternative explanation is possible. While considering that any target recruits an extended assembly of neurons, the generation of accurate interceptive saccades may result from the attenuation of the synaptic weight of spikes emitted by the cells that were the earliest excited by the passage of the target, combined with the promotion of the synaptic weight of spikes emitted by the latest and most recently excited cells.

The dorsolateral pontine nucleus (DLPN) is one of the nuclei that relays signals from motion-related areas of the cerebral cortex to the cerebellum. When a chemical lesion is placed in the DLPN, saccades toward a moving target are hypometric whereas saccades to a static target remain unchanged (May et al., 1988; Ono et al., 2003). The hypometria suggests that the saccades were directed toward the past location of the target. Neurons in DLPN project to the floccular and medio-posterior regions of the cerebellum. Ablation of the floccular region (flocculus and paraflocculus) impairs the velocity of slow eye movements, seemingly without affecting catch-up saccades (Zee et al., 1981). By contrast, unilateral inactivation of the output nucleus of the medio-posterior cerebellum, the caudal fastigial nucleus, impairs both contralesional saccadic and pursuit eye movements (Bourrelly et al., 2018a; 2018b). Because of the hypometria of saccades and the slowing of pursuit velocity, gaze direction always lags behind the target when it moves in the direction contralateral to the lesion.

< Figure 9 near here >

Concerning the generation of slow pursuit eye movements, the figure 9 schematizes the complex neuronal network involved in their generation when the movement direction is horizontal. The schema is not complete for it does not show how neurons in the oculomotor vermis and the rostral superior colliculi are being recruited during the generation of tracking (saccadic and slow) eye movements (Bourrelly et al., 2018b).

Some cells firing tonically in the nucleus prepositus hypoglossi (NPH) and the nearby medial vestibular nucleus (MVN) increase their firing rate during ipsilateral pursuit (McFarland and Fuchs, 1992). Although called "eye-head velocity" neurons in the monkey, we shall name them tonic neurons for simplicity and for distinguishing them from the burst neurons and EVN-1 described in the previous paragraphs. Some of the tonic neurons are believed to provide an excitatory input to the abducens neurons for contracting the agonist muscle fibers during ipsilateral pursuit. While documented in the cat, this ipsilateral excitatory tonic input (ETN) has been questioned in the monkey (Langer et al., 1986). Most NPH/MVN neurons seem to be inhibitory and to project to the contralateral side (synapses a) (McFarland and Fuchs, 1992; McCrea and Horn, 2006). When the sustained firing rate of tonic neurons is considered, we understand that the reciprocal commissural inhibition between the NPH/MVN in the left and right sides yields a push-pull organization, such that a reduced firing rate on one side enhances the firing rate on the opposite side. Thus, by suppressing the activity of the neurons that innervate the antagonist muscle fibers (red synapses a), the inhibitory tonic neurons (ITN) in the NPH/MVN contribute to the relaxation of these muscles, while disinhibiting the ipsilateral abducens and internuclear neurons (blue synapses a) and promoting an ipsiversive movement of both eyes.

Neurons in NPH/MVN are monosynaptically inhibited (synapse b) by Purkinje cells in the flocculus and the ventral paraflocculus (Lisberger et al., 1994), which receive excitatory input from the three major pontine nuclei (synapses c) involved in relaying visual motion signals to the cerebellum: the rostral part of nucleus reticularis tegmenti pontis (rNRTP), the dorsolateral pontine nucleus (DLPN) and the dorsomedian pontine nucleus (DMPN). These pontine nuclei contribute to the inhibition that Purkinje cells exert upon NPH/MVN neurons. After a chemical lesion in the dorsolateral pontine nucleus (DLPN) or in the rostral nucleus reticularis tegmenti pontis (rNRTP), ipsilesional pursuit eye movements are severely impaired (May et al., 1988; Ono et al., 2003; Suzuki et al., 1999). The suppression of pontine input likely reduces the floccullar inhibition of ITN neurons, which in turn impede the recruitment of agonist motor and internuclear neurons. The consequences of suppressing the input from DMPN have not been tested yet. However, an impairment of ipsilesional pursuit is expected because the DMPN and the contralateral caudal fastigial nucleus exhibit strong reciprocal connections, and inactivation of the latter severely impairs the generation of contralesional pursuit eye movements (Bourrelly et al., 2018b).

The NPH/MVN are also the target of signals from neurons in the pretectal nucleus of the optic tract (NOT; synapse d) (Mustari et al., 1994), some of which are sensitive to the motion of a small visual target (Fuchs et al., 1992; Mustari and Fuchs, 1990). These neurons fire during ipsilateral target motions (Mustari et al., 1990) and their loss severely impairs the ability to generate ipsilesional slow pursuit eye movements (Yakushin et al., 2000). NOT receives visual signals from the retina and from the middle temporal (MT) and medial superior temporal (MST) areas of cerebral cortex (synapse f) wherein numerous cells are sensitive to target motion as well (Distler et al., 2002). Projections of the frontal eye field (FEF) to NOT and to NPH have been documented (synapses e and e', respectively) (Distler et al., 2002). This access of frontal cortex to premotor neurons does not indicate a voluntary control of pursuit eye movements since the presence of a target is required for initiating a smooth pursuit. The NOT contributes to the maintenance of ipsilateral pursuit eye movements by its direct projections to tonic neurons in the ipsilateral NPH/MVN (synapse d), but also to the ipsilateral pontine nuclei (synapses g).

Through their projections to the ipsilateral rostral SC (synapse j), Gaba-ergic cells in NOT may contribute to suppression of the generation of saccades made in the direction opposite to the target motion (Büttner-Ennever et al., 1996). Simultaneously, they may depress the excitatory drive that NRTP exerts upon the ipsilateral paraflocculus (synapse k).

Finally, the secondary excitatory vestibular neurons (EVN-1 or PVP cells) that are involved in moving the eyes in the direction opposite to the head movement (Fig. 6) increase their firing rate during contraversive pursuit eye movements. However, as they do not seem to be inhibited by Purkinje cells located in the floccular region, the origin of the discharge modulation during contraversive pursuit remains to be identified. Direct projections from the frontal eye fields (FEF) to the NPH/MVN might support this influence.

## Conclusion

Studying orienting movements of gaze is a convenient means for investigating how intertwined networks of neurons in the central nervous system transforms sensory signals into motor commands, but also, how it enables an entire animal to visually locate and capture an object in the physical environment (Scudder et al. 2002; Goffart, 2017; Sparks, 2002) and to adapt when intrinsic damage or contextual changes alter their execution (Soetedjo et al., 2019).

During the last decades, neurophysiologists and neuroanatomists gathered in the monkey considerable knowledge that enabled identifying the core networks involved in the generation of eye movements (Büttner-Ennever, 2006; Fuchs et al., 1985; Goffart et al., 2017a; Grandin et al., 2002; Henn et al., 1984; Horn and Leigh, 2011; Krauzlis, 2004; Krauzlis et al., 2017; Moschovakis et al., 1996; Raphan and Cohen, 1978; Scudder et al., 2002; Sparks, 2002; Sparks and Mays, 1990; Ugolini et al., 2006). Technical developments offered the possibility to measure precisely the time course of these movements and to study correlations between the firing rate of neurons and kinematic parameters such as movement amplitude, velocity, acceleration and various differences (e.g., gaze error, velocity error, etc.). However, these correlation studies should not lead us to believe that a one-to-one correspondence exists between the multiple networks within which neuronal activities propagate, and the extrinsic and homogeneous medium (mathematical space) with which we quantify the motion of rigid objects such as the eyeballs, the head and targets. Such a mapping is not at all obvious; it is even questionable. Contrary to the physical space, the medium of neuronal activity is neither homogeneous nor passive. Unlike most objects at which we look and that we manipulate, the corresponding brain activity is not rigid. Even the mere spot of activity that a small static object evokes on the retina yields multiple parallel flows of activity that make its correspondence in the brain spatially distributed, temporally extended and context-dependent (Goffart et al., 2017a). Replacing neurons or neuronal chains to units or modules encoding geometric or kinematic relations between gaze and target conceals the muscle forces and the multiple antagonisms that we have described between the extraocular muscles and the recruited neuronal channels. Furthermore, the velocity profile of the movement and the time course of force development do not match (Miller and Robins, 1992). During horizontal saccades for example, the duration of the interval during which the muscle force increases is on average 60% shorter than the duration of the total change in eye orientation. A force decay happens well before the change in eye position ends and the larger the saccade, the longer its lead-time relative to saccade arrest (Lennerstrand et al., 1993). Finally, the temporal series of numerical values quantifying the changes during eye reorientation hides the antagonisms between those multiple channels that we have described. More crucially, for a given central neuron, the convergence of its presynaptic input from sources which are not only located in multiple neuronal groups but also conveyed with potentially diverse conduction speeds, entails that the sequence of spikes does not match anymore with the sequence of measured kinematic or kinetic values. As we move from the motor periphery to more central brain regions, the distributiveness of afferent input inevitably abolishes the correspondence between the temporal vicinity of successive spikes and the temporal vicinity of eye movement-related values.

In the majority of models proposed during the last decades, the movements of the eyes and the head were considered as driven by (intrinsic) error signals encoding displacement vectors in physical space (e.g., Lisberger et al., 1987). These models have been useful to communicate and to share a comprehensive picture of the complexity underlying the generation of movements. However, embedding geometric and kinematic notions within the inner functioning of the brain (i.e., mapping intrinsic neuronal signals with extrinsic behavioral measurements) may be neurophysiologically misleading because different constraints characterize the neurophysiological and kinematic descriptions. In fact, rather than outcomes of processes reducing geometric or kinematic errors, the saccade and pursuit deceleration may resume to transitions from imbalance to equilibria opposing populations of neurons whose activity leads to mutually antagonist movement tendencies (Goffart, 2019).

Further empirical investigations are required to determine and to explain several other issues. Indeed, numerous questions still remain unanswered such as: whether and how the networks underlying orienting movements of the eyes and head interact with those generating other types of goal-directed action like hand reaching movements or whole-body locomotion; whether and how they interact with the networks involved in the navigation and the memory of locations; whether and how they support the learning of new skills and possibly the acquisition of more abstract knowledge such as geometry or counting; etc. With the recent multiplication of cognitive studies that use eye-tracking techniques to explore with quantitative methods the so-called "inner space" (Johansson et al., 2013; Strohmaier et al., 2020), more effort is required to neurophysiologically characterize what are those covert processes that eye movements would express and how they connect to the core neuronal networks documented in our synthesis. Neurophysiological and neuroanatomical works must also continue to reveal the innervation of extraocular muscles compartments and unravel the multiplicity of channels and neuronal networks that enable a living organism to interact visually with its natural and social environment.

By studying the simple and seemingly straightforward problem of orienting the eyes toward the location of a physical object, we see how the neurophysiological and neuroanatomical studies performed with animals enables us to understand the symptoms exhibited by patients suffering from eye movement disorders (Horn and Leigh, 2011; Rucker and Lavin, 2021; Sharpe, 2008; Sharpe and Wong, 2005). They also show that the evolutionary and embryological developments led to patterns that are much more complex and redundant than the straightforward strategies of human engineers. Contrary to the case in which a design precedes the realization of a machine for solving a specific problem, Physiology teaches us that any behavioral observation is the macroscopic outcome of a multiple networks and parallel chains of microscopic elements embedded within the inner milieu. It also teaches us how the multiplicity and the diversity of elements enable an organism to solve several problems in parallel and even take up new challenges.

## Acknowledgements

The author is grateful to Drs Anja Horn, Richard J. Krauzlis, Paul J. May, Adonis Moschovakis, David L. Sparks and Catherine Vignal-Clermont for their corrections and suggestions to improve this article.

This work was possible thanks to support from Centre National de la Recherche Scientifique and from the Fondation pour la Recherche Médicale.

# Figure captions

**Figure 1**: Schematic representation of the extraocular muscles (A) and pulling directions engaged by the contraction of their fibers when the eyes are centered in the orbit (B). LR: lateral rectus, MR: medial rectus, SR: superior rectus, IR: inferior rectus, SO: superior oblique, IO: inferior oblique, tr: trochlea. The red arrows indicate the muscles whose contraction abducts the eye, the black arrows the muscles whose contraction adducts the eye. The turquoise arrows indicate the muscles whose contraction elevates the eye (supraduction), the yellow arrows those whose contraction infraducts the eye. The blue arrows indicate the muscles whose contraction causes an incyclotorsion (incycloduction) of the eye, the green arrows those whose contraction causes an excyclotorsion (excycloduction).

**Figure 2**: Neuronal network involved during horizontal saccades toward the left (A) and the right (B). Connecting lines ended by an arrow indicate excitatory connections; those ended by a circle indicate inhibitory synaptic connections. Blue color indicates the agonist neuronal elements, red color the antagonist ones. The thickness of connecting lines schematizes the strength with which the neurons fire. OPN: omnipause neurons, EBN: excitatory burst neurons, IBN: inhibitory burst neurons, ABD: abducens nucleus, OMN: oculomotor nucleus, MN: motoneurons, AIN: abducens internuclear neurons. At the bottom, the thickness of arrows attached to the eyeballs schematizes the strength of muscle contraction. LR: lateral rectus, MR: medial rectus, SR: superior rectus, IR: inferior rectus, SO: superior oblique, IO: inferior oblique. Between the panels A and B, a parasagittal section of the brainstem and cerebellum shows the approximate locations of the oculomotor nucleus (OMN), the paramedian pontine reticular formation (PPRF), the abducens nucleus (ABD) and the dorsal paragigantocellularis reticular formation (dPGRF). See text for explanations.

**Figure 3**: Neuronal network involved during vertical saccades toward the top (A) and the bottom (B). Connecting lines ended by an arrow indicate excitatory connections; those ended by a circle indicate inhibitory synaptic connections. Blue color indicates the agonist neuronal elements, red color the antagonist ones. The thickness of connecting lines schematizes the strength with which the neurons fire. uEBN: upward excitatory burst neurons, dEBN: downward excitatory burst neurons, uIBN: upward inhibitory burst neurons, TRO: trochlear nucleus, OMN: oculomotor nucleus, MN: motoneurons. On top, the thickness of arrows attached to the eyeballs schematizes the strength of muscle contraction. LR: lateral rectus, MR: medial rectus, SR: superior rectus, IR: inferior rectus, SO: superior oblique, IO: inferior oblique. Between the panels A and B, a parasagittal section of the brainstem and cerebellum shows the approximate locations of the rostral interstitial nucleus of the medial longitudinal fasciculus (RIMLF), the interstitial nucleus of Cajal (INC), the oculomotor nucleus (OMN) and the trochlear nucleus (TRO). See text for explanations.

**Figure 4**: Time course of a horizontal gaze shift toward a static visual target. A coil attached to the eye and to the head enabled to record their orientation in space. The movement of the eye in the orbit (blue trace) is obtained by subtracting the head orientation from the gaze orientation. The time interval during which the eye moves in the direction opposite to the head movement corresponds to the vestibulo-ocular reflex.

**Figure 5**: Neuronal network involved during combined horizontal eye and head movements toward the left. Connecting lines ended by an arrow indicate excitatory connections; those ended by a circle indicate inhibitory synaptic connections. Blue color indicates the agonist neuronal oculomotor elements, red color the antagonist ones. Green color indicates the cephalomotor elements. The thickness of connecting lines schematizes the strength with which the neurons fire. dSC: deep superior colliculus, RSN: reticulospinal neurons, EBN: excitatory burst neurons, IBN: inhibitory burst neurons, ABD: abducens motor and internuclear neurons, NMN: neck motoneurons, PVN: primary vestibular neurons, IVN: inhibitory vestibular neurons, EVN: excitatory vestibular neurons, VIII: eight cranial nerve.

**Figure 6**: Neuronal network involved in ending a leftward gaze shift while the head terminates its rotation (stabilized gaze direction). Connecting lines ended by an arrow indicate excitatory connections; those ended by a circle indicate inhibitory synaptic connections. Blue color indicates the agonist neuronal oculomotor elements, red color the antagonist ones. Green color indicates the neurons of the ascending tract of Deiters (ATDN). The thickness of connecting lines schematizes the strength with which the neurons fire. LR: lateral rectus, MR: medial rectus, ABD: abducens nucleus, OMN: oculomotor nucleus, MRMN: motor neurons innervating the medial rectus muscle, LRMN: motor neurons innervating the lateral rectus muscle, AIN: abducens internuclear neurons, NMN: neck motoneurons, PVN: primary vestibular neurons, IVN: inhibitory vestibular neurons, EVN: excitatory vestibular neurons, VIII: eight cranial nerve.

**Figure 7**: Neuronal network involved in ending a downward (A) and upward (B) gaze shift while the head terminates its movement (stabilized gaze direction). Connecting lines ended by an arrow indicate excitatory connections; those ended by a circle indicate inhibitory synaptic connections. Blue color indicates the agonist neuronal oculomotor elements, red color the antagonist ones. OMN: oculomotor nucleus, TRO: trochlear nucleus, IOMN: motor neurons innervating the inferior oblique muscle, IRMN: motor neurons innervating the inferior rectus muscle, SRMN: motor neurons innervating the superior rectus muscle, SOMN: motor neurons innervating the superior oblique muscle, PVN: primary vestibular neurons, IVN: inhibitory vestibular neurons, EVN: excitatory vestibular neurons, VIII: eight cranial nerve.

**Figure 8**: Neuronal network involved in the near response. LR: lateral rectus, MR: medial rectus, SP: sphincter pupillae, DP: dilator pupillae, CM: ciliary muscle, CG: ciliary ganglion, MRMN: motor neurons in the oculomotor nucleus that innervate the medial rectus muscle, EWpg: Edinger-Westphal paraganglionic nucleus, cMRF: central mesencephalic reticular formation, SOA: supraoculomotor area, cFN: caudal fastigial nucleus, SC: superior colliculus.

**Figure 9**: Neuronal network involved in the generation of horizontal slow pursuit eye movements. LR: lateral rectus, MR: medial rectus, ABD: abducens nucleus, OMN: oculomotor nucleus, MN: motor neurons, AIN: abducens internuclear neurons, NHP/MV: complex nucleus prepositus hypoglossi-medial vestibular nucleus, ETN: excitatory tonic neurons, ITN: inhibitory tonic neurons, cFN: caudal fastigial nucleus, NOT: nucleus of the optic tract, rSC: rostral superior colliculus, rNRTP: rostral part of nucleus reticularis tegmenti pontis, DMPN: dorsomedian pontine nucleus, DLPN: dorsolateral pontine nucleus, VPL-Th: ventroposterolaral part of thalamus, FEF: frontal eye field, MST: medial superior temporal area of cerebral cortex.

# Figure 1

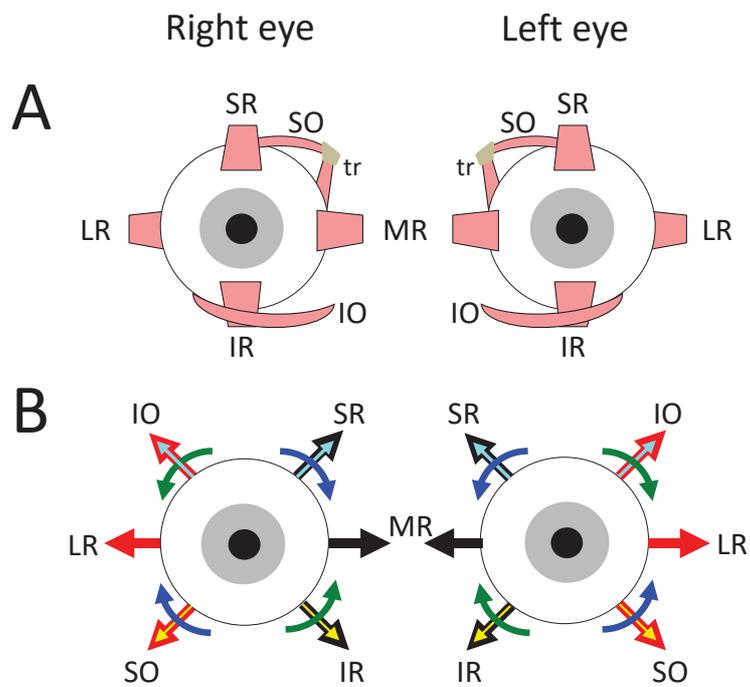

# Figure 2

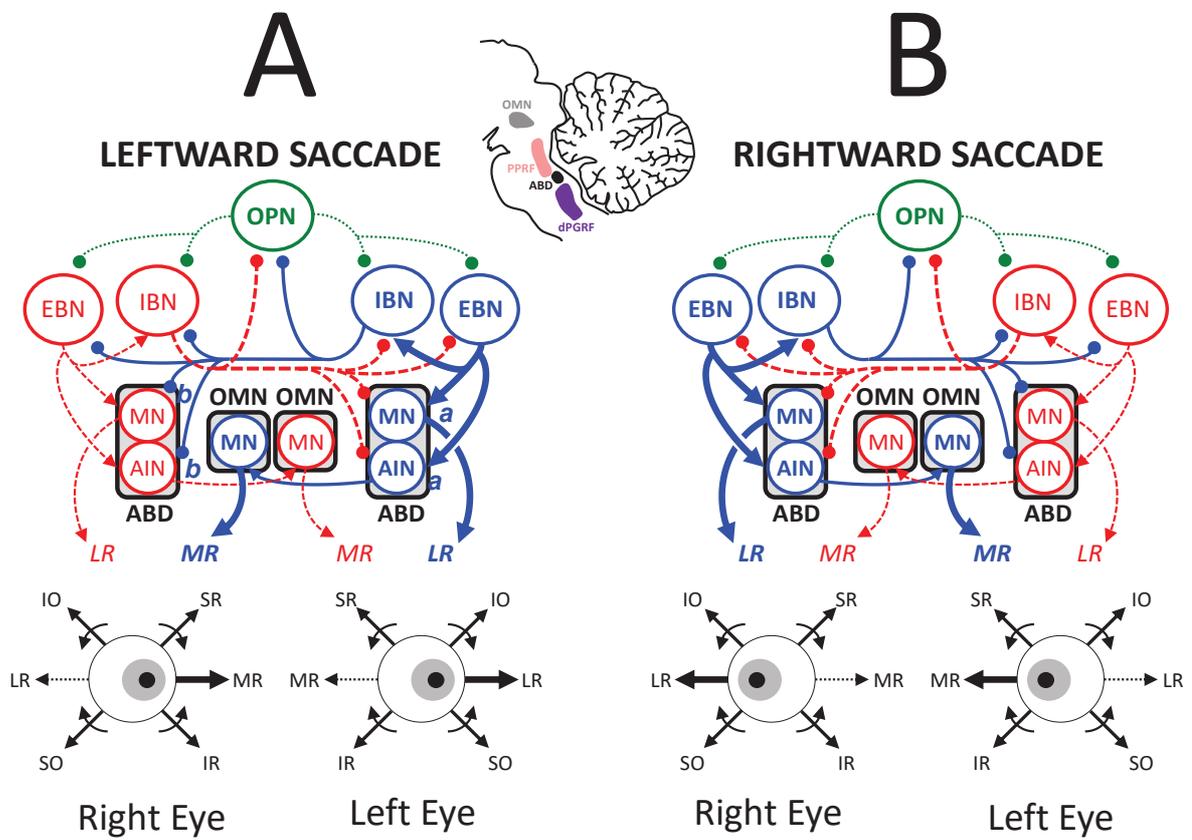

# Figure 3

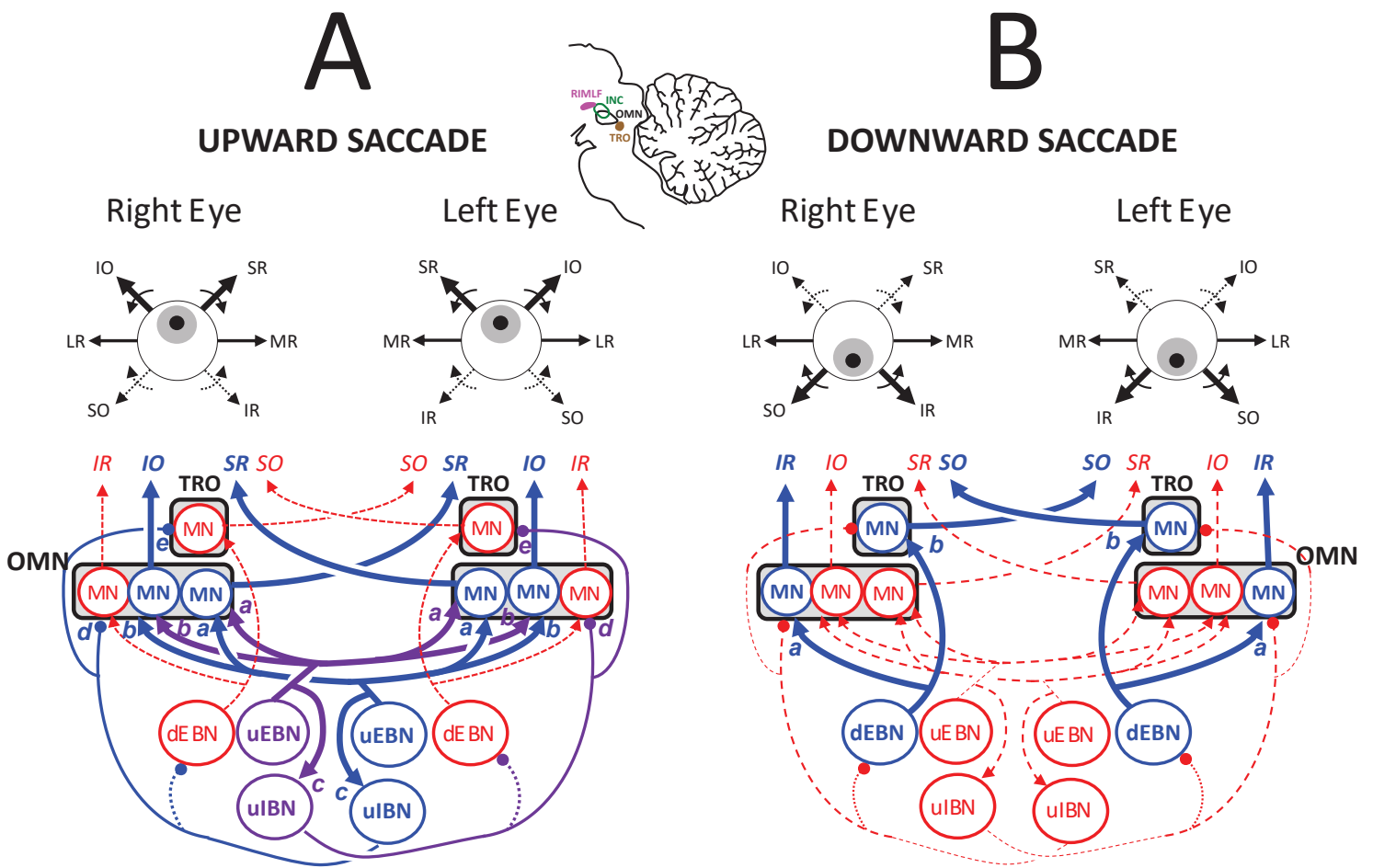

# Figure 4

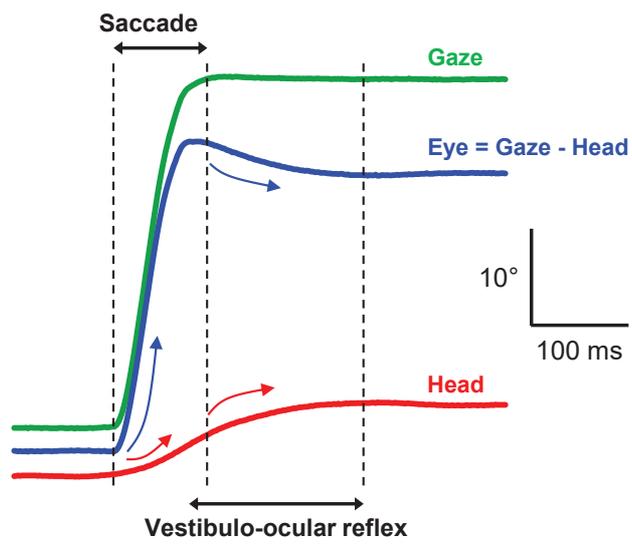

# Figure 5

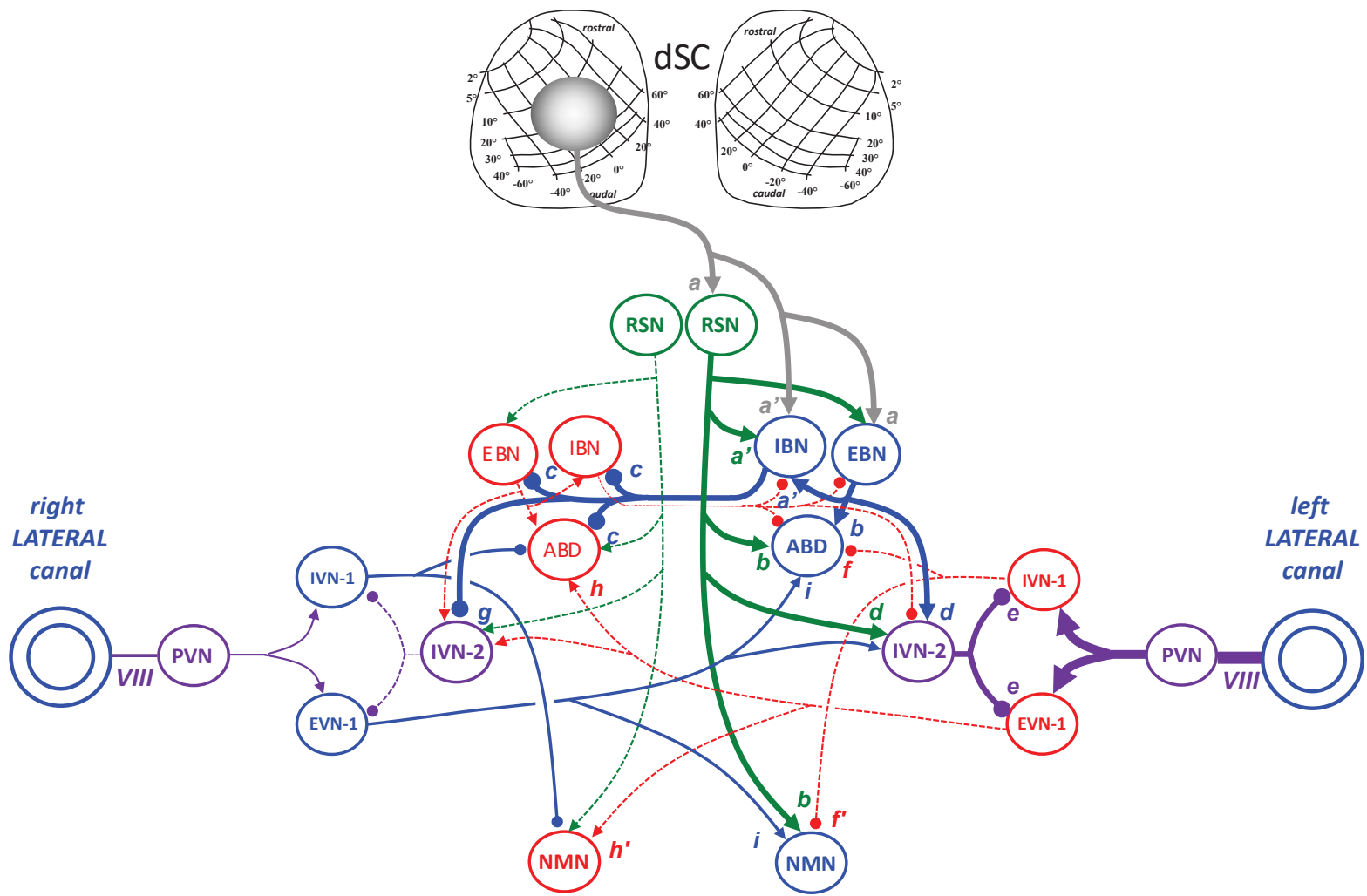

# Figure 6

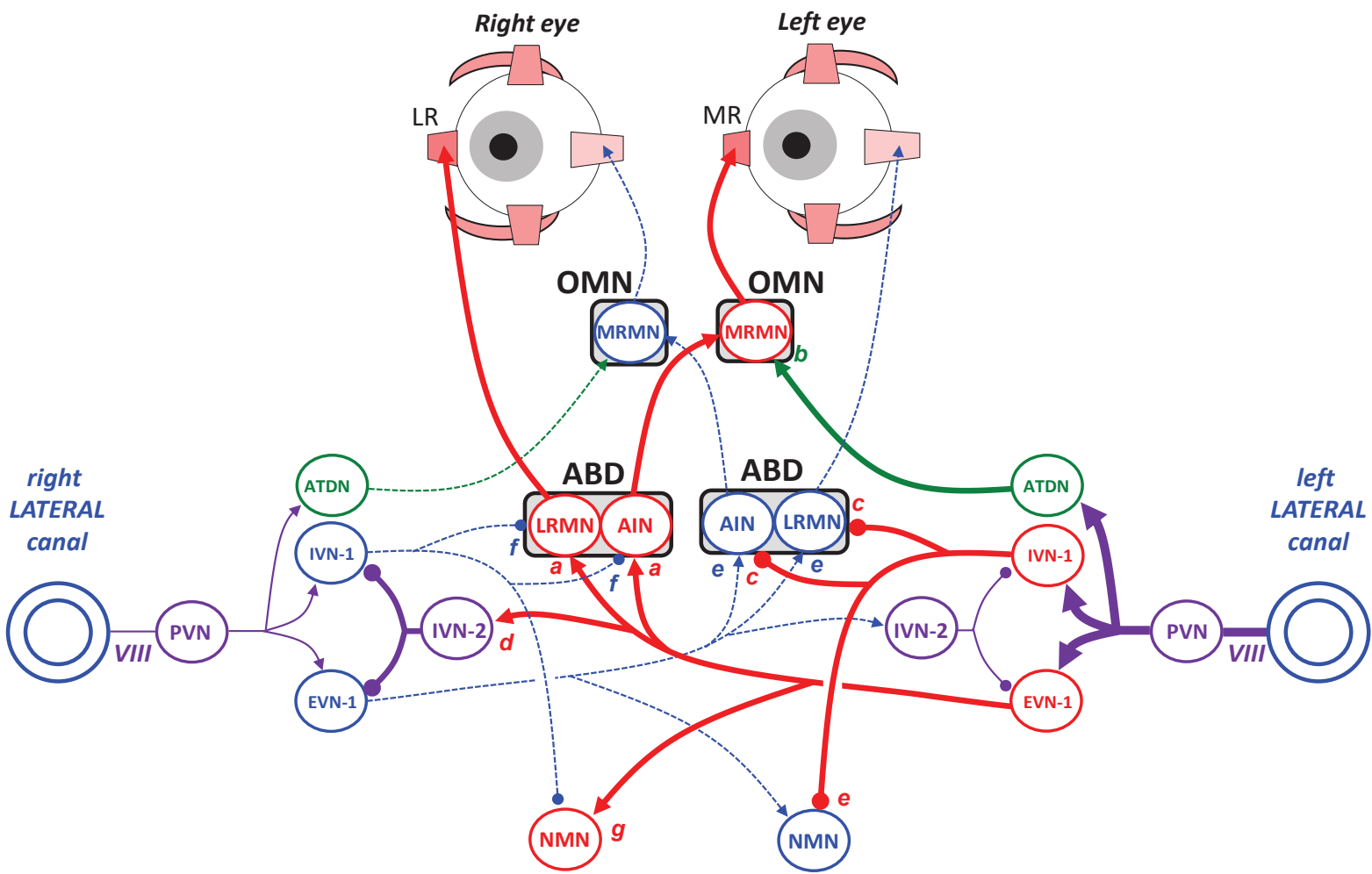

# Figure 7

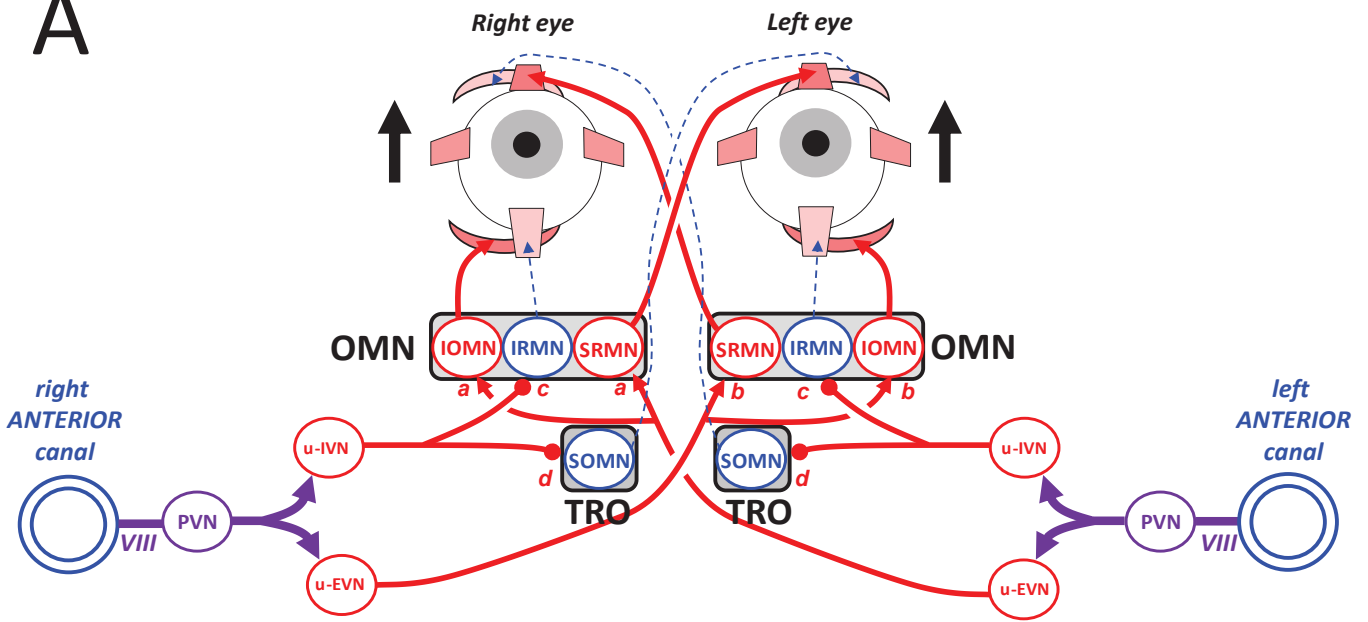

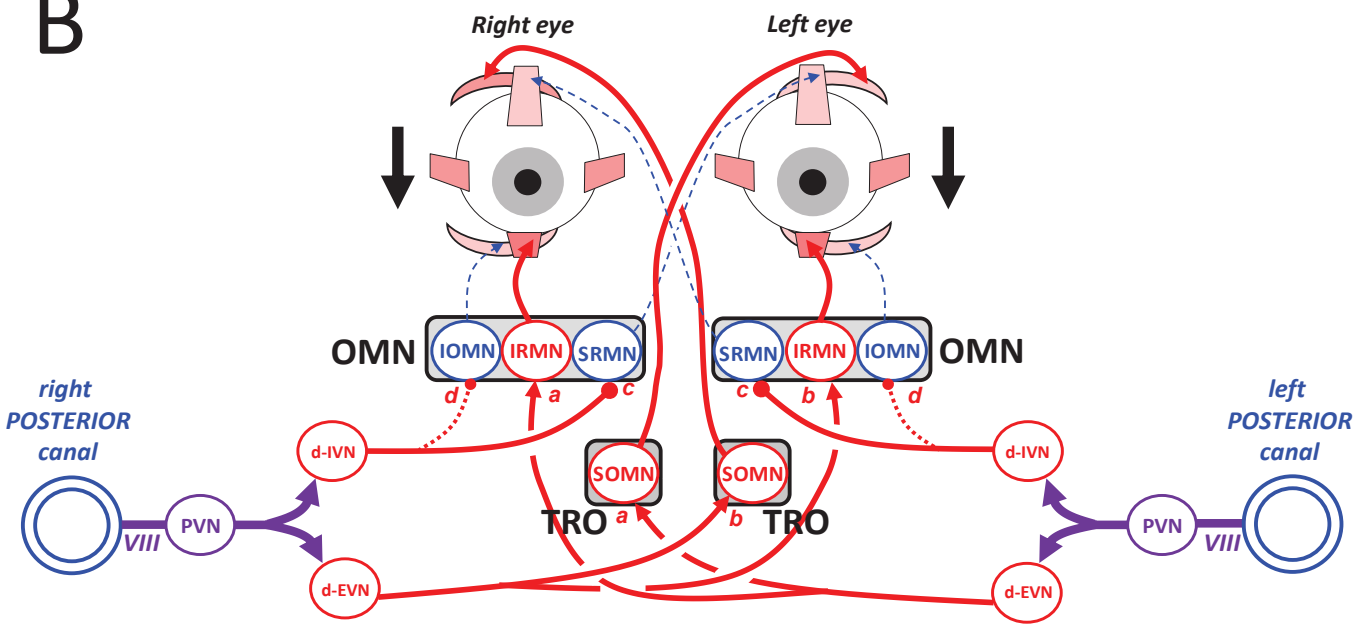

# Figure 8

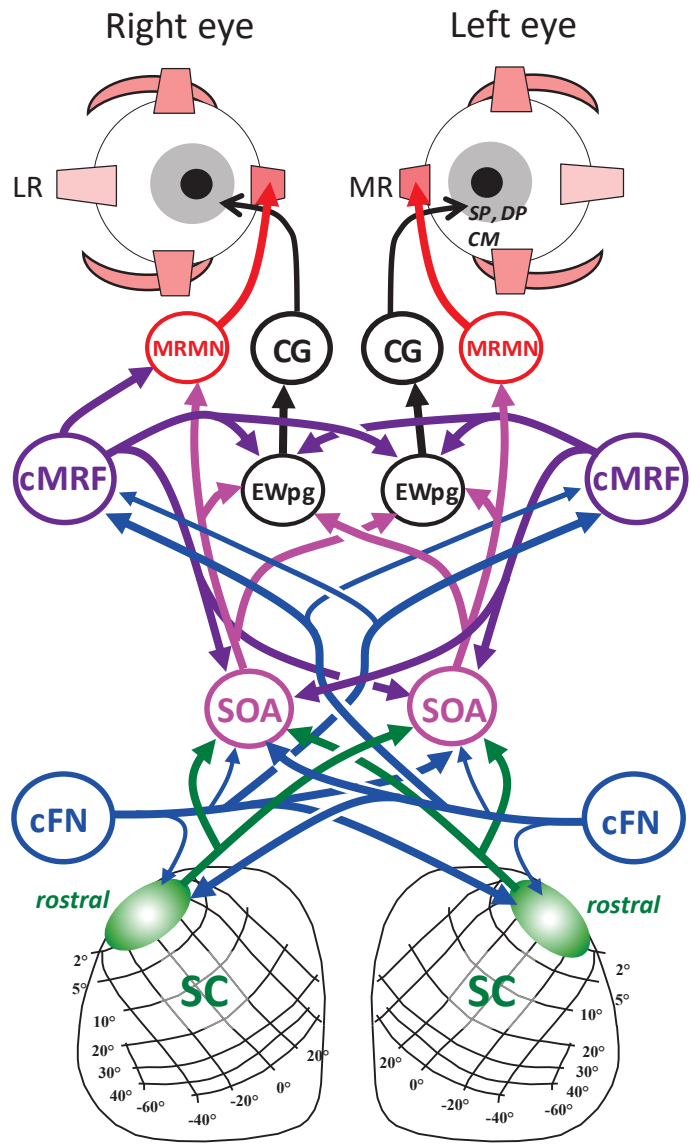

# Figure 9

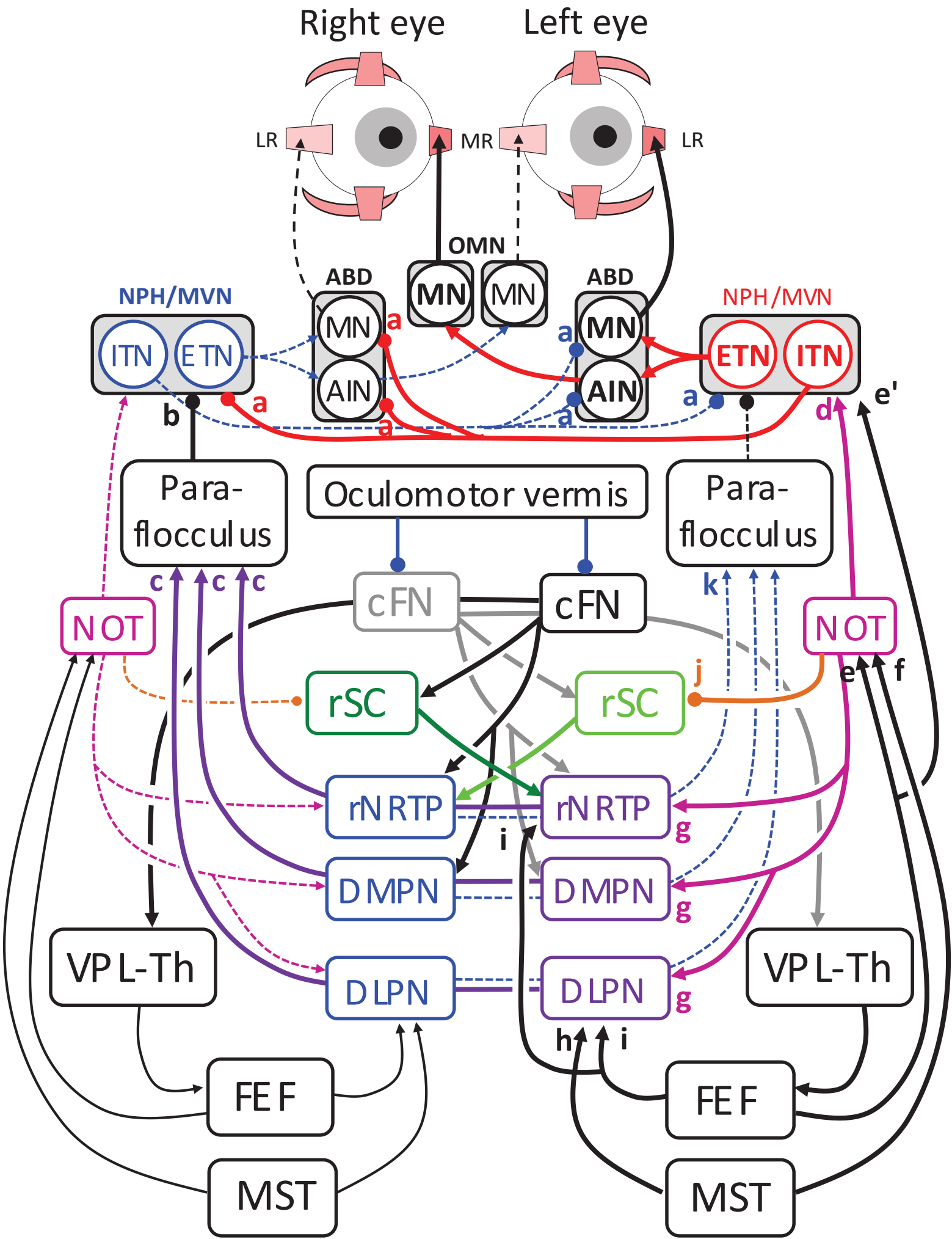